\documentclass[conference]{IEEEtran}
\IEEEoverridecommandlockouts
\usepackage{cite}
\usepackage{comment}
\usepackage{pgfplots}
\usepackage{amsmath,amssymb,amsfonts}
\usepackage{algorithmic}
\usepackage{subfig}
\usepackage{graphicx}
\usepackage{caption}
\usepackage{textcomp}
\usepackage{xcolor}
\usepackage{hyperref}
\hypersetup{colorlinks,allcolors=black}

\def\BibTeX{{\rm B\kern-.05em{\sc i\kern-.025em b}\kern-.08em
    T\kern-.1667em\lower.7ex\hbox{E}\kern-.125emX}}
    
    \newcommand{\RNum}[1]{\uppercase\expandafter{\romannumeral #1\relax}}

\makeatletter
    \newcommand{\linebreakand}{%
      \end{@IEEEauthorhalign}
      \hfill\mbox{}\par
      \mbox{}\hfill\begin{@IEEEauthorhalign}
    }
    \makeatother
\begin{document}

\title{A DPU Solution for Container Overlay Networks
\thanks{Laboratory under Contract No. DE-AC02-05CH11231 with the U.S. Department of Energy. The U.S. Government retains, and the publisher, by accepting the article for publication, acknowledges, that the U.S. Government retains a non-exclusive, paid-up, irrevocable, world-wide license to publish or reproduce the published form of this manuscript, or allow others to do so, for U.S. Government purposes.}
}

\author{\IEEEauthorblockN{Anton Njavro}
\IEEEauthorblockA{\textit{Department of Computer Science} \\
\textit{Boston University}\\
Boston, MA\\
njavro@bu.edu}
\and
\IEEEauthorblockN{James Tau}
\IEEEauthorblockA{\textit{NVIDIA} \\
Westborough, MA \\
jtau@nvidia.com}
\and
\IEEEauthorblockN{Taylor Groves}
\IEEEauthorblockA{\textit{NERSC} \\
\textit{Lawrence Berkeley National Lab}\\
Berkeley, CA \\
tgroves@lbl.gov}
\linebreakand
\IEEEauthorblockN{Nicholas J. Wright}
\IEEEauthorblockA{\textit{NERSC} \\
\textit{Lawrence Berkeley National Lab}\\
Berkeley, CA \\
njwright@lbl.gov}
\and
\IEEEauthorblockN{Richard West}
\IEEEauthorblockA{\textit{Department of Computer Science} \\
\textit{Boston University}\\
Boston, MA \\
richwest@bu.edu}
}

\maketitle
\begin{abstract}
There is an increasing demand to incorporate hybrid environments as part of workflows across edge, cloud, and HPC systems. In a such converging environment of cloud and HPC, containers are starting to play a more prominent role, bringing their networking infrastructure along with them. However, the current body of work shows that container overlay networks, which are often used to connect containers across physical hosts, are ill-suited for the HPC environment. They tend to impose significant overhead and noise, resulting in degraded performance and disturbance to co-processes on the same host.

This paper focuses on utilizing a novel class of hardware, Data Processing Unit, to offload the networking stack of overlay networks away from the host onto the DPU. We intend to show that such ancillary offload is possible and that it will result in decreased overhead on host nodes which in turn will improve the performance of running processes.
\end{abstract}


%%%This contains the submitted extended abstract
%%%New edits should be in the individual files (e.g. intro.tex)

\section{Introduction}
\label{sec:intro}
There is an increasing demand to incorporate hybrid environments as part of complex workflows across edge, cloud and HPC systems.  This is driven by the desire to connect together sensors, elastic resources and cloud-native frameworks with tightly coupled, high-performance systems to create workflows between scientific instruments, geographically distributed computing and computationally intensive simulations~\cite{bard2022lbnl, osti_1769755}.

\begin{figure}
    \centering
    \includegraphics[width=0.99\linewidth]{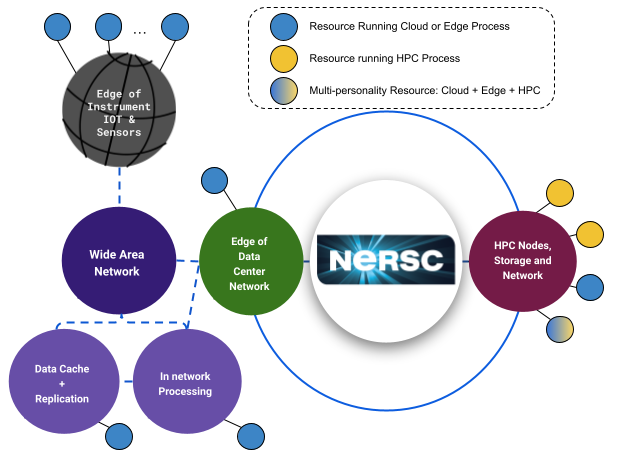}
    \caption{Vision of converged HPC, Cloud and Edge workflow.  HPC resources will be responsible for supporting both HPC and cloud/edge-native applications and software.}
    \label{fig:workflow}
\end{figure}

For an example of this vision, consider Figure~\ref{fig:workflow}.  In this Figure, cloud and edge processes may run throughout the entire span of the figure, from the instrument, wide-area network (WAN), data caches, data center and HPC system.  These processes may run on dedicated HPC resources (solid blue) or may exist as a service (such as analytics or machine learning framework), alongside HPC simulations on a shared node (gradient blue-yellow).

These cloud or edge processes regularly use TCP/IP networking and containers to provide portability across a diverse computing platforms.
Current container networking architectures predominantly rely on overlay networks in order to connect containers across different physical hosts while providing isolation. However, these container overlay networks (and software-defined networking) can be resource intensive to the performance detriment of co-processes residing on the same node\cite{firestone2018azure}. 
Results from Suo et al.~\cite{10.1145/3409334.3452040} show that overlay networks might impose up to 3.5X context switches on the host system, causing a decrease in performance for container application along with other co-processes running on the host machine.

In this paper, we consider how we can better support cloud-native overlay networks so that they may be deployed alongside existing HPC workloads, many of which show substantial sensitivity to system-noise~\cite{petrini2003case}. We do this by exploring the capabilities of the Nvidia Bluefield 2 Data Processing Unit (DPU)~\cite{NvidiaBF2} to provide ancillary offload~\cite{groves2021use} for Docker~\cite{docker} overlay networks.

Our paper makes the following contributions:
\begin{itemize}
    \item Characterization of approaches to offload aspects of cloud-based container networking.
    \item Offloading the networking stack of container overlay networks onto the DPU.
    \item Measuring the performance benefits of ancillary offload of overlay networks.
\end{itemize}

The rest of this paper is structured as follows: section \RNum{2} introduces existing body of work analyzing overlay networks and their detrimental impact on host performance. It then introduces the concept of Data Processing Unit, with special focus on characteristics of NVIDIA's BlueField 2. Section \RNum{3} introduces our design and covers implementation details. Section \RNum{4} measures performance evaluation of DPU solution against conventional Docker overlay network and host-to-host networking. We conclude with section \RNum{5} where we discuss future direction of our research.
\section{Background and Related Work}
\label{sec:background}
\subsection{Container Overlay Networks}
\begin{figure*}
    \centering
    \includegraphics[width=1\linewidth]{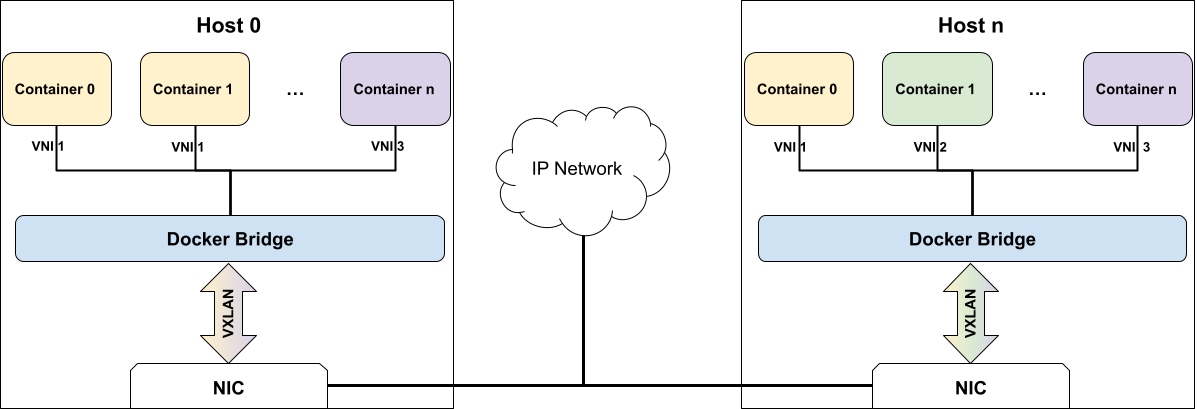}
    \caption{Example Docker overlay network.  Several containers sharing a single host and network card utilize VXLAN to provide isolation by providing OSI layer 2 Ethernet encapsulation over OSI layer 3 IP networks.  This allows for composing flexible network partitions across network boundaries.  In this example Container-0 and Container-1 on Host-0 share a virtual layer-2 network (VNI 1) with Container-0 on Host-n.}
    \label{fig:currentoverlay}
\end{figure*}
%need to describe what a container overlay network is and why we need them.
OSI Layer-2 (data link) overlay networks are a common abstraction for partitioning and connecting containers across different host machines. This is accomplished by attaching virtual network identifiers to enclosing protocols, which allow switches to restrict and direct traffic accordingly
There are many flavors of technologies that enable layer-2 overlay networks this such as, Virtual Local Area Network (VLAN)~\cite{581577}, Virtual Extensible LAN (VXLAN) \cite{Mahalingam2014VirtualEL}, NVGRE~\cite{Garg2015NVGRENV}, GENEVE~\cite{rfc8926}.  For the purposes of this work we utilize VXLAN.  A full survey of these technologies is beyond the scope of this work.  
The isolation often comes at the cost of performance as it increases the depth and complexity of the networking stack.
An example of the Docker\cite{docker} overlay networking architecture can be seen in Figure~\ref{fig:currentoverlay}.  In this example several containers on two different hosts are partitioned into a separate layer-2 virtual networks using VXLAN. (As indicated by virtual network identifier number -- VNI.)  Specifically, on Host-0, Container-0 and Container-1 share VNI-1 with Container-0 of Host-n.  The layer 2 virtual overlay spans across the layer-3 IP network.
There exists a current body of work focused on container networks, their performance analysis, and ways to improve on performance~\cite{8485865}\cite{10.1145/3409334.3452040}\cite{227669}. Research focused on overlay networks in particular has shown that they suffer from significant performance degradation caused by a large number of software interrupts combined with load imbalance~\cite{10.1145/3409334.3452040}.

%SRIOV, VF, PF, VXLAN, Representor

\subsection{Open Virtual Switch}
Open Virtual Switch is an open-source virtual switch used in Linux, it was created with primary purpose of connecting hypervisors and virtual machines to the outside networks and between themselves~\cite{10.5555/2789770.2789779}.
It relies on advanced flow classification and caching techniques in order to achieve higher throughput performance with reduced overhead imposed on hypervisor or host system.  Furthermore, OVS provides a means for separating VXLAN-based flows across containers on the same host. OVS design also allows for greater flexibility when it comes to network design which is a crucial aspect of software-defined Networking (SDN). Two main components of OVS are the \textit{ovs-vswitchd} daemon running in user space and the kernel datapath. The main idea is that the first packet of a flow creates a miss, resulting in the kernel module forwarding the packet up to the \textit{ovs-vswitchd} daemon in user space. The user space daemon then caches the forwarding decision before forwarding the packet.  Future packets of the same flow do not require the ovs-vswitchd lookup in user space.

NVIDIA also provides its ConnectX-6 network interface cards with Accelerated Switching And Packet Processing ($ASAP^2$)\cite{asapp} technology which allows for OVS hardware offloading by moving the OVS data plane on ConnectX-6 hardware while not altering the control plane.  This means initial packets will take the path through ovs-vswitchd in userspace, but later packets of a flow are processed on the NIC.  Such offload results in better performance allowing users to handle high-throughput networks with reduced overhead. 

\subsection{Virtual Interfaces}

Virtual machines and containers needing to share hardware resources attached to a node over PCIe (such as a network adapter) may use single-root IO virtualization (SR-IOV) to expose the PCIe resources and provide isolation.
This enables containers and virtual machines to bypass hypervisors and software switches that would otherwise perform memory and interrupt translations.
In SR-IOV there are concepts of virtual functions (VF), which allow the representation of a virtual network adapter.  Each VF is associated with a physical function (PF) which represents an actual physical device.

BlueField 2 uses the netdev representor model\cite{representorsnetdev} that is used with SR-IOV to map host-side physical and virtual functions on the DPU. Representors serve as a tunnel in order to pass traffic  to adequate physical or virtual function. They are also used to set up embedded switch on ConnectX-6 for the corresponding interfaces. Our approach creates one VF representor for every container created on the host side. We then connect those representors to the OVS bridge running on the DPU Arm cores in order to correctly forward the traffic to the appropriate containers.

\subsection{Data Processing Unit (DPU)}

Overhead of high-throughput networking, resulting in degraded host performance, has been a growing issue in recent years facing both cloud computing and scientific computing domains. Therefore, it was not surprising to see that some of the first solutions for offloading that overhead of high-throughput networks onto the specialized hardware came from cloud providers themselves \cite{firestone2018azure} \cite{AWSNitro}. Initial solutions were primarily focused on an FPGA-based approach with highly specialized offload domains. However, both cloud providers and the research community started to notice the possibility of a more programmable approach \cite{8850757} in which SmartNICs could be leveraged to build software-defined networking. NVIDIA entered the market with the purpose of combining general programmability and performance, resulting in the release of their line of BlueField SmartNICs. As their design and approach stood out significantly from the previous generations of SmartNICs, with their ability to combine compute and network capabilities, they introduced new terminology for such hardware, calling it a data processing unit (DPU)\cite{NVIDIADPU}. 

BlueField 2 DPU is a programmable processing element consisting of a tightly coupled network adapter, eight 64-bit Armv8 A72 cores, and various hardware accelerators. Current DPUs come in a variety of forms including FPGA, multi-core, and many-core architectures~\cite{groves2021use}.  While prior work has examined offloading portions of software-defined networking stack within data centers to FPGA-based DPUs~\cite{firestone2018azure}, to our knowledge no existing work has explored whether CPU-based DPUs are sufficient to allow substantial reduction of the noise cloud-native containerization can create for co-resident HPC applications in an edge-to-HPC type workflow. Performance characteristics of BlueField 2 have been studied \cite{liu2021performance}, and they show that BlueField 2 holds potential for offloading. It is important to note that while BlueField 2 itself might be underpowered as compared to current host-server systems, primarily due to weaker Arm A72 cores, BlueField 2 still has potential for improving host performance through ancillary offload. This host versus DPU performance difference should be reduced with the upcoming line of BlueField 3 DPUs.

Bluefield 2 has four modes of operation. $1)$ \textit{\textbf{Embedded Function/DPU mode}}: DPU mode is the default mode BlueField 2 operates in. In DPU mode all the resources of the DPU are controlled by the Arm cores. Additionally, all the networking traverses through a virtual switch sitting on those Arm cores. This is the mode used to offload the overlay networking stack in our approach.
$2)$ \textit{\textbf{Restricted mode}}: Restricted mode is a more secure extension of DPU mode in which administrators are not allowed to access the DPU through the host. Its primary purpose is for supporting the zero-trust security principles.
$3)$ \textit{\textbf{NIC mode}}: In NIC mode DPU acts as a regular network interface card. It is still accessible through host for firmware updates, but the traffic does not go through the Arm cores.
$4)$ \textit{\textbf{Separated Host mode}}: Separated Host mode allows both the host and the DPU to receive traffic simultaneously. They have distinct MAC addresses, and can have distinct IP addresses, allowing for DPU to host communication.

BlueField 2 comes with support on the software side through NVIDIA DOCA SDK framework\cite{DOCA}. DOCA framework serves as an interface through which developers can fully utilize hardware capabilities of DPUs and conjoin it with their applications running on host server. DOCA also allows for easier development experience and comes with additional benefits such as backwards compatibility, which will allow developers to reap benefits of improved hardware in the upcoming BlueField 3 DPU. In that sense DOCA seems to be for DPUs what CUDA \cite{CUDA} is for GPUs.
\subsection{Default Approach to Docker Networking}
\label{sec:defaultapp}
Docker overlay networks are a primary option for creating distributed networks across multiple hosts with Docker daemon. Being overlay networks, they use encapsulation and decapsulation in order to abstract the networking structure underneath them. As shown in figure \ref{fig:currentoverlay}, Docker overlay networking stack consists of two main components on the host system. $1$) \textbf{\textit{VXLAN interface}} on the client side is used to encapsulate outgoing packets with 50 byte VXLAN headers, while on the server side it serves to decapsulate respective packets. One benefit of VXLAN is that it removes the need for L2 connectivity between hosts since VXLAN does the tunneling, allowing for greater flexibility while designing the network. $2$) \textbf{\textit{Docker bridge}} is a software bridge used to connect all the containers residing on the same host. Docker bridge also serves as a gateway to those containers connecting them to the outside network. Bridge relies on VLAN in order to deliver packets to containers, and is coordinated through the Docker swarm services.
Container service management traffic and data traffic run on the same network, however service management traffic is encrypted. While not by default, data can be encrypted as well using the AES\cite{117146} algorithm.
\section{Proposed Solution}
\label{sec:solution}

Container overlay networks tend to impose high overhead on host systems primarily due to the multilayered path a packet has to take between a container and host physical interface\cite{10.1145/3409334.3452040}. As a part of this network path, overlay networks utilize virtual bridges and VXLAN tunnels, both of which may cause significant noise in terms of interrupts and context switches. Our solution focuses on the idea of ancillary offload for overlay networks using the DPU. 

We propose a design approach in which VXLAN encapsulation and decapsulation, along with bridging of multiple containers are moved onto the DPU. We replace the Linux bridge on the host machine with an OVS bridge on the DPU and make use of ConnectX-6 VXLAN offload capabilities. Single Root I/O Virtualization (SR-IOV), more specifically a custom SR-IOV plugin for containers \cite{SRIOVPLUGIN}, is used in order to give each container access to high-throughput network interface. Development was done on two types of machines, one being 32-core Intel Xeon server, and another server being Dell R7525 equipped with AMD EPYC processor running on CloudLab \cite{Duplyakin+:ATC19} infrastructure.

\subsection{DPU offload}
Our design is focused on moving host-based bridge and VXLAN processing onto the DPU while maintaining isolation between containers by utilizing SR-IOV.  This approach reduces CPU overhead on the host and allows us to potentially utilize BlueField 2 hardware accelerators for high-performance custom overlay solutions.

One of the issues we faced in our approach was the way in which we would connect containers residing on the host to this network. Since the central part of our overlay network resided in a different hardware domain than our containers did, it was not possible to encompass them within one namespace. A couple of bridging techniques were investigated as possible solutions for this problem. The first one was to connect VF interfaces on the host machine to containers using Docker MACVLAN networking mode. In such mode, the network driver assigns MAC address to the container’s virtual network interface, which in our case would be VF on the host side. MACVLAN would allow us to explicitly assign VFs to container namespaces. However, MACVLAN networking comes with multiple pitfalls, some of them being security and portability. MACVLAN is also discouraged in the cloud industry with some vendors such as AWS not supporting it. Another approach we considered was using a bridge networking driver. In such a scenario, all the containers on the host would attach to one bridge which would then in turn attach to PF. The issue of design redundancy came up, as we were already using an OVS bridge residing on the DPU. Such an approach while easier to implement and port added unneeded noise and design repetition. Therefore, we decided to use the Mellanox SR-IOV container plugin which allowed us to automatically activate VFs and connect them to the appropriate containers. In addition, once we have properly configured VF representors on the DPU side our solution persisted. Two main benefits of the SR-IOV plugin are ease of implementation, since it automatically handles bringing up VFs and assigning them directly to each container, and tight-coupling with Mellanox/NVIDIA hardware. The plugin does however have an issue of not being actively supported, and it has been succeeded by the SR-IOV Network Device Plugin for Kubernetes \cite{kubernetes}.

\begin{figure}
\centering
\includegraphics[width=1\linewidth]{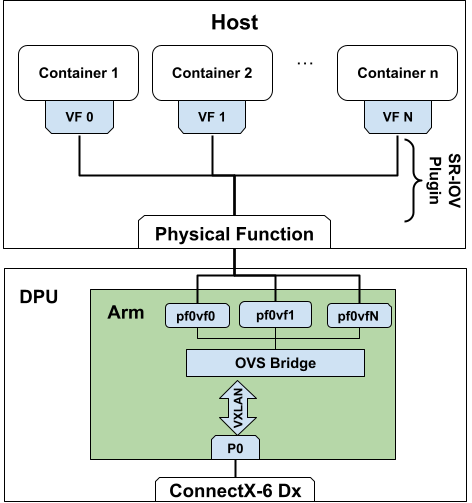}
\label{fig:Ng1} 
\caption{Container overlay network solutions}
\end{figure}

\section{Evaluation}
\label{sec:eval}
\subsection{Experiment Setup and Configuration}
As mentioned in the previous section, we ran the development and testing of our system on two kinds of servers. Initial development was done on CloudLab machines, more specifically their Dell R7525 nodes situated at the Clemson cluster. Our setup used two of those nodes, one serving as a client and the other as a server node. R7525 server comes equipped with two 32-core AMD EPYC processors, 512GB of RAM, two NVIDIA GPUs, and BlueField 2 DPU. Nodes were connected with 100Gb Ethernet connection allowing us to test our approach on networks seen in high-performance cloud environments.
The software environment running on our CloudLab R7525 machine was Ubuntu 20.04 with kernel version 5.4.0. Docker version was 20.10.18 community edition.

After development and testing on R7525 nodes, we noticed that an issue regarding SR-IOV Virtual Functions (VFs) occurred. We observed a sharp and unexpected decrease in throughput through VFs for which we haven't found a cause or a solution yet. We did test the same design on two different servers, one containing Intel Xeon CPU, and another containing AMD CPU. We observed expected throughput through VFs in both scenarios.

To measure throughput we used \textit{iPerf}\cite{iperf} in order to send packets with 8KB buffer length between the server and client node. We also ran \textit{iPerf} within containers in order to measure performance it achieves through overlay network. Additionally we used \textit{Perf}\cite{perf} and \textit{mpstat} offered by Linux to measure system statistics. 

Our primary goal was to determine whether the ancillary offload of the overlay networking stack contributed to reduced disturbances on the host system. Therefore our metrics of interest were: the number of context switches, number of interrupts generated, total throughput, and percent of CPU time left for the user.

\begin{figure}[htbp]
\centering
\subfloat[Percentages of available CPU cycles]{\label{fig:a}\includegraphics[width=0.75\linewidth]{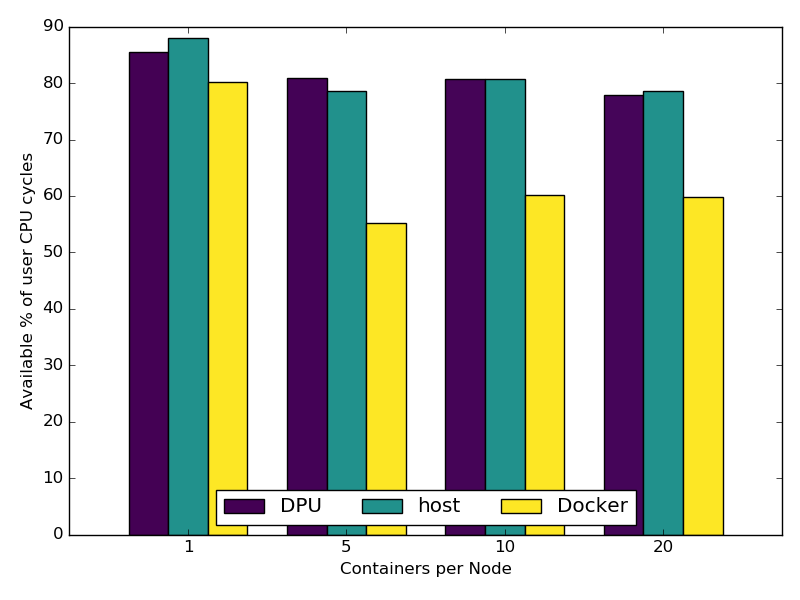}}\qquad
\subfloat[Rate of interrupts per throughput]{\label{fig:b}\includegraphics[width=0.75\linewidth]{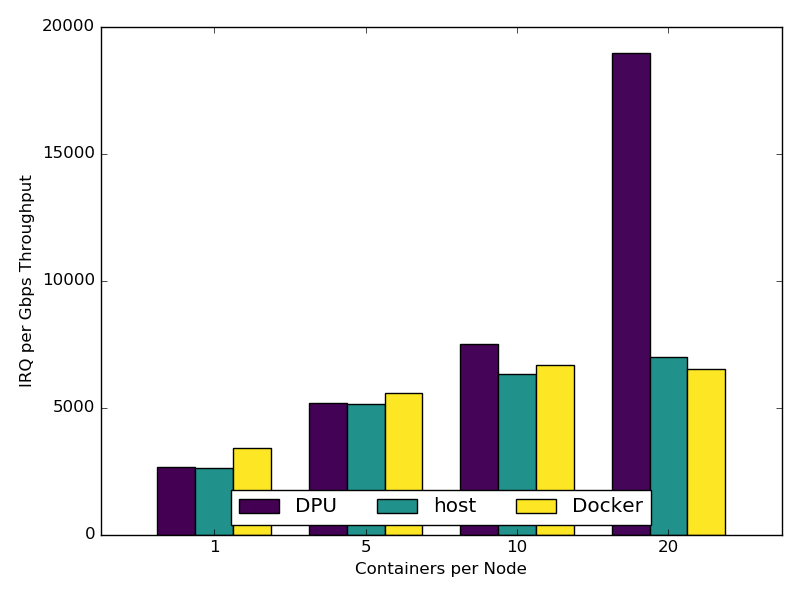}}\\
\subfloat[Total networking throughput]{\label{fig:c}\includegraphics[width=0.4\textwidth]{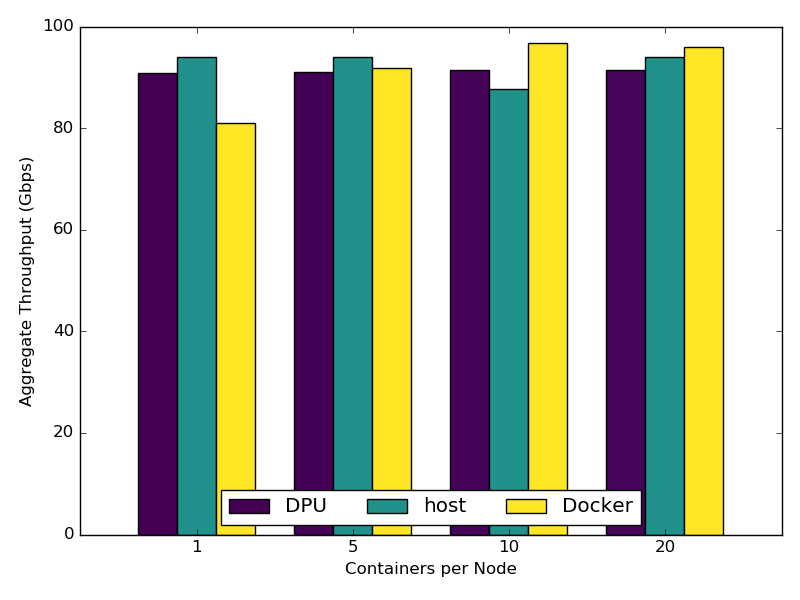}}\qquad%
\subfloat[Rate of context switches per throughput]{\label{fig:d}\includegraphics[width=0.4\textwidth]{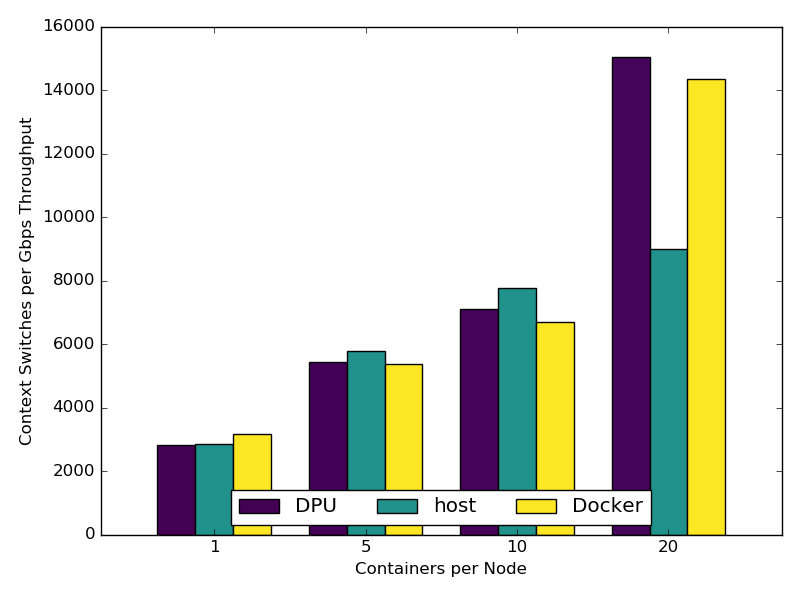}}%
\caption{Performance measurements}
\label{fig:performance measurements}
\end{figure}

\subsection{Throughput}
As figure \ref{fig:performance measurements} c) shows, throughput in all three cases is relatively high averaging around 90 Gb/s. In the event of a single container running on the host the default overlay network, with a throughput of 80.95 Gb/s, does not lag significantly behind host-to-host and DPU solutions. As we increase the number of containers on the host we see that all of them seem to successfully handle high throughput, actually noticing increases. While we did not test our system with varying packet sizes there is a possibility of decreased performance on smaller packet sizes \cite{10.1145/3409334.3452040}. Additionally, we observed unusual throughput with an increased number of containers especially pertaining to the Docker overlay network. As can be seen from the charts Docker overlay network seems to perform better than the host-to-host setup which is unexpected and requires further analysis in future research.

\subsection{Context Switches and Interrupts}
Another metric of our interest was context switches as they tend to approximate the amount of overhead and disturbance created for the host system and co-assigned processes. Our charts in fig. \ref{fig:performance measurements} show number of context switches normalized for throughput. For one container we observe as expected that the default overlay network has a higher amount of context switches, even though not as high as we have seen in some prior research on the subject. As the number of containers increases to 5 and 10 we see a steady rise of generated context switches, however at 20 containers we see a significant jump for both the DPU solution and the default overlay network. Similar unexpected behavior occurs for interrupts where we see a large spike for our DPU approach with 20 containers. While the detailed benchmark analysis of networking internals is out of the scope of this paper, these phenomena will be studied further in our future research. Context switches and interrupts were measured using \textit{Perf} and \textit{mpstat} respectively.

\subsection{CPU Utilization}
To measure user CPU utilization, we first spawned enough \textit{stress}\cite{stress} program threads to keep every core occupied with user-level work. Since \textit{stress} runs purely in the user space, and since our system did not have any background programs running, we know that almost 100\% of CPU time should be dedicated to the user. Therefore, any significant drop in user CPU time will signify overhead imposed by the system.
User CPU utilization is one of the main metrics we were interested in regard to our DPU offload approach since the main objective of ancillary offload was to reduce system overhead imposed by the default Docker overlay network. Figure \ref{fig:performance measurements} a) does show that our design with DPU offload allows for notably more CPU time to be dedicated to the user, hence imposing a lower overhead. It is also interesting to note that while the default overlay network observes a larger drop in user CPU time going from one to five containers, it seems to stabilize and even rise slightly with 10 to 20 containers. It is also interesting that the number of context switches and interrupts does not seem to correspond with the available CPU time to the user. Again, our future work will focus on a more detailed analysis of these discrepancies accompanied with real-world workloads running on host as opposed to only synthetic benchmarks.

%\pgfplotsset{compat=1.17}

\section{Conclusions and Future Work}
\label{sec:conclusions}

As we have seen, profiling container networks introduces novel challenges when measuring an increasingly large set of containers. Our work proposes the idea of utilizing a novel class of hardware, a Data Processing Unit, to offload the default Docker overlay networks which are currently fully situated on the host. Our proof-of-concept of ancillary offload allows for a higher amount of CPU time dedicated to the user. However, we do observe that the issue of context switches and interrupts does persist. We suspect that the SR-IOV container plugin might be at fault, as it contains certain SR-IOV devices which are not included in the Default overlay network. Since the plugin is not currently supported by NVIDIA we recommend using a more novel Kubernetes SR-IOV plugin for which there is an active support. While our design showcases the simple ancillary offload benefits of DPU, new contributions that fully utilize the capabilities of various hardware accelerators situated on the BlueField 2 will reap the benefits of more custom and nuanced networking designs without the high system burden imposed on the host. Our performance analysis also showed a discrepancy in performance overheads of Docker overlay networks observed in Suo et al. \cite{10.1145/3409334.3452040} paper. We suspect that the difference arises due to two main reasons. First, Suo et al. paper uses two virtual machines on one physical node to represent host-to-host networking while we use two physical nodes without any virtual machine overhead. Second, our approach measures systems with 1,2,5,10,20 client/server pairs of containers on each physical node while Suo et al. paper focuses on one pair of client/server containers. Future research should also focus on analyzing overlay networking stack from the perspective of a larger number of containers running in parallel.   

Our future work will encompass more detailed profiling of the overlay networking stack combined with the DPU offload. We also expect to analyze the upcoming BlueField 3 DPU and utilize its even larger set of hardware accelerators to implement more advanced network designs.

\bibliographystyle{IEEEtran}
\bibliography{refs}

\end{document}